\documentclass[prl,aps,twocolumn,preprintnumbers,showpacs]{revtex4}

\begin{document}
\title{Operator-sum representation of time-dependent density operators and its applications}
\author{D. M. Tong$^{1,2}$, L. C. Kwek$^{1,3}$,
C. H. Oh$^1$, Jing-Ling Chen$^1$, and L. Ma$^1$ }
\address{$^1$Department of Physics, National University of
Singapore, 10 Kent Ridge Crescent, Singapore 119260, Singapore \\
$^2$Department of Physics, Shandong Normal University, Jinan 250014, People's Republic of China\\
$^3$ National Institute of Education, Nanyang Technological
University, 1 Nanyang Walk, Singapore 639798, Singapore }
\date{\today}

\begin{abstract}
We show that any arbitrary time-dependent density operator of an
open system can always be described in terms of an operator-sum representation (Kraus
representation) regardless of its initial condition and the path
of its evolution in the state space, and we provide a general
expression of Kraus operators for arbitrary time-dependent density
operator of an $N$-dimensional system. Moreover, applications of
our result are illustrated through several examples.
\end{abstract}
 \maketitle

Arbitrary states of any quantum system, be it open or closed, can
always be described by density matrices $\rho(t)$. If the system
is closed, its time evolution is unitary, and there exists a
unitary operator $U(t)$, such that
$\rho(t)=U(t)\rho(0)U(t)^\dagger$, where $\rho(0)$ is the initial
state. However, if the system is open, the evolution is not
necessarily unitary and the above relation between $\rho(t)$ and
$\rho(0)$ is in general not valid. To describe the evolution of
an open system, one usually employs the Kraus
representation\cite{Kraus}. If there exist operators $M_\mu(t)$,
called Kraus operators, that satisfy
\begin{eqnarray}
\rho(t)=\sum\limits_\mu M_\mu(t)\rho(0)M_\mu(t)^\dagger,
\label{rhot}
\end{eqnarray}and
\begin{eqnarray}
\sum\limits_\mu M_\mu(t)^\dagger M_\mu(t)=I, \label{mm}
\end{eqnarray}
then the evolution of $\rho(t)$ is said to have operator-sum
representation or the Kraus representation.

It has been known that a completely positive map always possesses
the operator-sum representation while a general positive map may
not
do\cite{Kraus,Preskill,Pechukas,Pomero,Bouda,Philip,Gen,Peixoto,Mika,Sonja}.
The papers\cite{Peter, Salgado, Hayashi} furthermore pursued the
existence of the Kraus representation by investigating the role
of initial correlations between the open system and its
environment, and showed that a map based on the reduced dynamics,
in general, cannot be described as the Kraus representation in
the presence of the initial correlations because an additional
inhomogeneous part appears. On the other hand, we know that for a
qubit system,  although a non-completely positive map
$\$:\rho(0)\rightarrow\rho(t)$ does not possess the Kraus
representation with the Kraus operators independent of $\rho(0)$,
the state $\rho(t)=\$(\rho(0))$ can still be expressed in the
form of Eq. (\ref{rhot}) with the operators $M_\mu(t)$ selected
for $\rho(0)$\cite{Philip,Gen,Tong1}. Arbitrary density matrices
$\rho(t)$ of a qubit system can always be described in terms of
the operator-sum representation. This start us to wonder whether
the same property is valid for $N$-dimensional systems. That is,
for any arbitrary time-dependent state $\rho(t)$ and initial state
$\rho(0)$ of an $N$-dimensional system, is it true that the
operators $M_\mu(t) $ can always be found so that (\ref{rhot}) and
(\ref{mm}) are fulfilled? This is a interesting issue as it
extends the notion of the Kraus representation associated with a
map to that associated with an evolution of the state and we find
that it is useful to some physical problem.

In this paper, we prove that the time-dependent density operator
of an arbitrary $N$-dimensional quantum system can always be
described in terms of the Kraus representation and provide a
general expression of Kraus operators for the $N$-dimensional
system. Some applications of our result are also discussed.

Firstly, we prove that any arbitrary two density matrices $\rho_A$
and $\rho_B$ of an $N$-dimensional open system can always be
connected by Kraus operators.

Suppose
\begin{eqnarray}
\rho_A=\sum\limits_{i=1}^N p^A_i |A_i\rangle\langle A_i|=U_A\rho^d_AU_A^\dagger,\label{rhoab1}\\
\rho_B=\sum\limits_{i=1}^N p^B_i |B_i\rangle\langle
B_i|=U_B\rho^d_BU_B^\dagger, \label{rhoab2}
\end{eqnarray}
where $p^A_i$ ($p^B_i$) and $|A_i\rangle$ ($|B_i\rangle$) are the
eigenvalues and the orthonormal eigenvectors of the density
matrix $\rho_A$ ($\rho_B$) respectively, $\rho^d_A$ ($\rho^d_B$)
is the diagonal matrix with the entries $p^A_i$ ($p^B_i$), and
$U_A$ ($U_B$) is the unitary transformation matrix, the $i$-th
column of which is just the vector $|A_i\rangle $ ($|B_i\rangle$).
We want to prove that there exist the operators $M_\mu$,
satisfying
\begin{eqnarray}
&&\rho_B=\sum\limits_\mu M_\mu\rho_AM_\mu^\dagger,\label{rhoa}\\
&&\sum\limits_\mu M_\mu^\dagger M_\mu=I. \label{mm2}
\end{eqnarray}

To find the required Kraus operators, one may write $M_\mu$ as
$N\times N$ matrices with unknown elements and then directly
solve  (\ref{rhoa}) and (\ref{mm2}) to determine the matrices.
However, this method is not feasible for high dimensional systems
due to the computational complexity. To overcome the problem, we
first seek diagonal matrices, which are generally easier to
manipulate. To this end, we look for operators $M'_i$ such that
\begin{eqnarray}
&&\rho^d_B=\sum\limits_\mu M'_\mu \rho^d_A {M'_\mu}^\dagger,\label{rhod}\\
&&\sum\limits_\mu {M'_\mu}^\dagger M'_\mu=I. \label{mm3}
\end{eqnarray}
To solve the above equations, we obtain the matrix operators
$M'_\mu $ ($\mu=0,1,2,...N-1$) with the entries
\begin{eqnarray}
 (M'_\mu)_{ij}= \sqrt{p^B_i}\delta_{i,j-\mu}+\sqrt{p^B_i}\delta_{i,j-\mu+N}~~,
\label{m}
\end{eqnarray}
where $i,j=1,2,...N$\cite{Tong2}.

With the matrices $M'_\mu $, we can construct the Kraus operators $M_\mu $
satisfying (\ref{rhoa}) and (\ref{mm2}). To see this, we use the
relations,
\begin{eqnarray}
\rho^d_A=U_A^\dagger\rho_AU_A,\label{rhoab3}\\
\rho^d_B=U_B^\dagger\rho_BU_B. \label{rhoab4}
\end{eqnarray}
Substituting (\ref{rhoab3}) and (\ref{rhoab4}) into (\ref{rhod}),
we have
\begin{eqnarray}
U_B^\dagger\rho_BU_B=\sum\limits_\mu M'_\mu U_A^\dagger\rho_A
U_A{M'_\mu}^\dagger,
\end{eqnarray}
that is,
\begin{eqnarray}
\rho_B=\sum\limits_\mu U_B M'_\mu U_A^\dagger\rho_A (U_BM'_\mu
U_A^\dagger)^\dagger.
\end{eqnarray}
Let
\begin{eqnarray}
M_\mu=U_BM'_\mu U_A^\dagger, \label{mmu}
\end{eqnarray}
it is obvious that $M_\mu $ defined by (\ref{mmu}) satisfy
(\ref{rhoa}) and (\ref{mm2}). They are the Kraus operators
connecting $\rho_A$ with $\rho_B$.  Since $U_A$, $U_B$ and
$M'_\mu $ can be  deduced directly from $\rho_A$ and $\rho_B$,
the explicit expression of $M_\mu$ is obtained for arbitrary
density matrices.

Secondly, having proved that any arbitrary two density matrices
$\rho_A$ and $\rho_B$ of an $N$-dimensional open system can
always be connected by Kraus operators, we now revert to our
discussion concerning the time evolution of open systems. We show
that the time-dependent state of an arbitrary open system can
always be described in terms of the Kraus representation.

We replace $\rho_A$ and $\rho_B$ in the above demonstration by
$\rho(0)$ and $\rho(t)$ respectively. Note that for any
evolution of an $N$-dimensional open system, the state $\rho(t)$
and the initial state $\rho(0)$ can always be expressed as
\begin{eqnarray}
\rho(t)=\sum\limits_{i=1}^N p_i(t) |\psi_i(t)\rangle\langle
\psi_i(t)|
,\label{rhot1}\\
\rho(0)=\sum\limits_{i=1}^N p_i(0) |\psi_i(0)\rangle\langle
\psi_i(0)|.
\label{rhot2}
\end{eqnarray}
By comparing (\ref{rhot1}) (\ref{rhot2}) with (\ref{rhoab1})
(\ref{rhoab2}) and using (\ref{mmu}), one can immediately write
down the Kraus operators $M_\mu (t)$ as
\begin{eqnarray}
M_\mu(t)=U(t)M'_\mu(t) U(0)^\dagger, \label{mmt}
\end{eqnarray}
where $U(t)$ ($U(0)$) is the unitary transformation matrix
that diagonalizes $\rho(t)$ ($\rho(0)$) given explicitly by
\begin{eqnarray}
&U(t)&=\left(\begin{array}{ccccc}\psi_1(t)~&\psi_2(t)~&...&...&~\psi_N(t)
\end{array}\right)\label{ut},\\
&U(0)&=\left(\begin{array}{ccccc}\psi_1(0)~&\psi_2(0)~&...&...&~\psi_N(0)
\end{array}\right),
\label{u0}
\end{eqnarray}
and $M'_\mu(t) $ ($\mu=0,1,...N-1$) are given by
\begin{eqnarray}
&M'_0(t)&=\left(\begin{array}{ccccc}
\sqrt{p_1(t)}&0&0&...&0\\
0&\sqrt{p_2(t)}&0&...&0\\
0&0&\sqrt{p_3(t)}&...&0\\
 &...&...&...&\\
0&0&0&0&\sqrt{p_N(t)}
\end{array}\right),
\nonumber
\end{eqnarray}
\begin{eqnarray}
&M'_1(t)&=\left(\begin{array}{ccccc}
0&\sqrt{p_1(t)}&0&...&0\\
0&0&\sqrt{p_2(t)}&...&0\\
 &...&...&...&\\
0&0&0&...&\sqrt{p_{N-1}(t)}\\
\sqrt{p_N(t)}&0&0&0&0
\end{array}\right),\nonumber
\end{eqnarray}
\begin{eqnarray}
...~~~~~~~...~~~~~~~...~~~~~~~\nonumber
\end{eqnarray}

\begin{eqnarray}
&M'_{N-1}(t)&=\left(\begin{array}{ccccc}
0&0&...&0&\sqrt{p_1(t)}\\
\sqrt{p_2(t)}&0&0&...&0\\
0&\sqrt{p_3(t)}&0&...&0\\
 &...&...&...&\\
0&0&0&\sqrt{p_N(t)}&0
\end{array}\right).
\label{m0m1}
\end{eqnarray}
The operators $M_\mu (t)$ defined by (\ref{mmt})-(\ref{m0m1})
satisfy (\ref{rhot}) and (\ref{mm}), giving the Kraus
representation of $\rho(t)$. Generally speaking, one cannot assert
that any arbitrary time-dependent density operator possesses the
Kraus representation simply from the result that any two states
can be connected by Kraus operators. However, here we can make
the assertion because $p_i(t)$ are always positive and
(\ref{m0m1}) is always valid independent of time $t$.

So far, we have shown that the time-dependent state of an open
system can always be described in terms of the Kraus
representation. A general expression of Kraus operators for any
arbitrary $N$-dimensional system is given. When the evolution of
the open system is given by $\rho(t)$ and $\rho(0)$, the Kraus
operators can be written immediately as $M_\mu(t)=U(t)M'_\mu(t)
U(0)^\dagger$, where $U(t)$, $U(0)$ and $M'_\mu(t)$ are
explicitly given by (\ref{ut}), (\ref{u0}) and (\ref{m0m1}).
Since the Kraus operators are not unique, the expression
(\ref{mmt}) is only one set of them. The other equivalent
expressions of the Kraus operators can be obtained by
$\tilde{M}_\mu(t)=\sum\limits_\nu M_\nu(t) V_{\mu\nu}$, where
$V_{\mu\nu}$ are the elements of an arbitrary unitary matrix $V$.

Thirdly, in the following paragraphs, we will illustrate some applications of our result.

The result can help to clarify some ambiguous concepts. For
example, the {\it{Kraus representation theorem}}, recalling the
well-known {\it{representation theorem}}, which states that a map
has the Kraus representation if and only if it is linear,
completely positive (CP) and trace preserving\cite{Kraus}, one
may thought that the expression $\rho'=\$(\rho)$ has no Kraus
representation if the map $\$$ is not CP. However, our result
shows that any two states $\rho$ and $\rho'$ can always be
connected by the Kraus operators irrespective of the form of the
map acting on $\rho$. This shows that although non-CP map does
not possess the Kraus representation, the expression
$\rho'=\$(\rho)$ can still be cast into the Kraus representation.
For instance, the transposition operator $T: \rho \rightarrow
\rho^T$ is positive but not completely positive. According to the
{\it{Kraus representation theorem}}, there does not exist the
Kraus representation of the map $T$. However, the expression $
\rho^T=T(\rho)$ can still be described by Kraus operators. Let
$\rho_A=\rho$ and $\rho_B=\rho^T$, the Kraus operators can be
obtained directly using (\ref{mmu}). The important key for
clarifying the ambiguity is to note that there is a difference
between the Kraus representation of a map and the Kraus
representation for a state under the action of a map. The Kraus
operators describing the Kraus representation of a map are
independent of the state $\rho$ while the Kraus operators
describing the Kraus representation of a state $\rho$ under the
action of a map may be dependent on the state. A non-CP map has
no Kraus representation, but $\rho'=\$(\rho)$, be $\$$ a CP map
or not, always has the Kraus representation.

An important corollary of our result is that there always exists
a CP map between any two quantum states and the map can be
represented through $N$ Kraus operators. This corollary is
obvious, because any two states can be connected by $N$ Kraus
operators $M_\mu $ and the CP map can then be defined by the $N$
Kraus operators. The corollary shows that even for such two
states $\rho$ and $\rho'$, where $\rho'$ is obtained from $\rho$
by a non-CP map  $\$:\rho\rightarrow\rho'$, one can still find an
alternative map ${\tilde{\$}}$ that is completely positive,
satisfying ${\tilde{\$}}:\rho\rightarrow\rho'$.

The conclusion that an arbitrary time-dependent state $\rho(t)$
can always be described by the Kraus representation could have
other deep applications in the study of open systems, especially
in the field of quantum information. For example, it is useful
for the study on geometric phase. As we know, geometric phases of
both pure state and mixed state under unitary evolutions have been
clarified\cite{Berry,Aharonov,Samuel,Mukunda,Pati,Sjoqvistm,Tong}.
A new issue is on the geometric phases for open systems under
nonunitary evolutions. Some papers\cite{Peixoto,Ericsson,Carollo}
just use the Kraus operators $M_\mu$ to define and calculate the
geometric phases of open systems. As described in ref.
\cite{Ericsson}, the relative phases are defined by
$\alpha_\mu=\arg tr[M_\mu(\tau)\rho(0)]$, and the geometric
phases can be calculated by making polar decomposition of
$M_{\mu}(t)$, such that $M_{\mu}(t)=h_{\mu}(t)u_{\mu}(t)$, where
$h_\mu(t)$ are Hermitian and positive, and $u_\mu(t)$ are
unitary. The relative phases $\alpha_\mu$ will then lead to the
geometric phase when $u_{\mu}(t)$ satisfy the $N$ parallel
transport conditions $\langle \psi_i(0)|u_{\mu}(t)^+\dot
{u}_{\mu}(t)|\psi_i(0)\rangle =0,~~(i=1,2,...,N)$. With the
scheme showed in the present paper, any time-dependent density
matrix can be easily written as the Kraus representation. So one
can transplant the notion of geometric phases defined by using
Kraus operators to arbitrary evolutions of quantum systems by
writing $\rho(t)$ as the Kraus representation.

Another example of its applications concerns the inverse problem
of the evolutions of open systems. Suppose that the density
matrix  of a system is given as a time-dependent function
$\rho(t)$, one wants inversely to deduce the evolutional
operators or Hamiltonians that evolve the initial state $\rho(0)$
to the state $\rho(t)$. This is an important issue because
physicists sometimes need to prepare experimentally a quantum
system that is expected to evolve along a given path in the
projected Hilbert space. The approach for solving the problem is
to  constitute a closed system by combining the open system with
an ancilla. The open system will undergo nonunitary evolution
while the combined system evolves unitarily as
$\varrho(t)=U_{sa}(t)\varrho(0)U_{sa}(t)^+$, where $U_{sa}(t)$
are unitary operators acting on the combined system and
$\varrho(0)=\rho(0)\otimes|0_a\rangle\langle 0_a|$ is the initial
state. The issue becomes to finding the unitary operators
$U_{sa}(t)$ so that $tr_e\varrho(t)=\rho(t)$. It is quite a
difficult problem in general. However, the present paper can
provide an effective approach to obtain the unitary operators. As
concluded above, evolutions of an open system  always have the
Kraus representation and the Kraus operators can be directly
deduced by the given $\rho(t)$.  With the Kraus operators
$M_\mu(t)$, one can easily obtain the unitary operator
$U_{sa}(t)$.  In fact, in order to satisfy
$tr_e\varrho(t)=\rho(t)$, the elements of $U_{sa}(t)$ in the
bases $|\Psi_i(0)\rangle \otimes|j_a\rangle$ are only required to
be $[U_{sa}(t)]_{ij,k0}=[M_{j}(t)]_{ik}$ while
$[U_{sa}(t)]_{ij,kl}~(l\neq 0)$ are arbitrary but keeping
$U_{sa}(t)$ to be unitary and $U_{sa}(t)|_{t=0}=I$,  where
$|j_a\rangle~ (j=0,1,...K-1)$ are the bases of the ancilla.
Obviously, there are infinitely many such unitary operators
$U_{sa}(t)$, so do the Hamiltonians $H=i\dot U(t)U(t)^+$. One can
choose the suitable ones which can be easily performed in the
laboratory.

It is also interesting to note that the Kraus operators provided
in the present paper have some intriguing properties. Since these
Kraus operators are dependent on the eigenvectors of ``input
state" $\rho_A$ but independent of its eigenvalues, all the input
states with distinct eigenvalues but the same eigenvectors will
yield the same ``output" state $\rho_B$. That is, the map defined
by the Kraus operators can transform all the input states whose
Bloch vectors lie on  the same diameter of the Bloch sphere to
the same output state. Specially, one may let
$\rho_A=\frac{1}{N}I$, and then has $\rho_B=\frac{1}{N}\sum M_\mu
M_\mu^\dagger$. We suppose that these properties of the Kraus
operators may be useful in the study of open systems.

The work was supported by NUS Research Project Grant: Quantum
Entanglement with WBS R-144-000-089-112. J.L.C. acknowledges
financial support from Singapore Millennium.


\begin{thebibliography}{99}
\bibitem{Tong2} Noticing that $\sum_{\mu=0}^{N-1}(\delta_{i,k-\mu}+\delta_{i,k-\mu+N})(\delta_{j,k-\mu}+\delta_{j,k-\mu+N})= \delta_{ij}$ $(i,j,k=1,2,...,N)$, one can verify that Eq. (9) satisfies Eqs. (7) and (8) by showing $(\sum\limits_\mu M'_\mu \rho^d_A {M'_\mu}^\dagger)_{ij}=p^B_i\delta_{ij}$ and $(\sum\limits_\mu {M'_\mu}^\dagger M'_\mu)_{ij}=\delta_{ij}$.
Besides, Eq. (9) can be written as $M_\mu=\sqrt{\rho^d_B}S^\mu$, where $S$ is a shift unitarity defined by $S_{ij}=\delta_{i+1,j}+\delta_{N,j+N-1}$. With this form of $M_\mu$,  the verification can also be carried out by direct matrix operations.
\end{thebibliography}
\end{document}